\documentstyle[12pt]{article}
\title{\large\bf
REALITY CONDITIONS FOR ASHTEKAR VARIABLES
AS DIRAC CONSTRAINTS
}
\author{
Hugo A. Morales--T\'ecotl$^{1,3}$\thanks{hugo@xanum.uam.mx},
        Luis F. Urrutia$^{1,2}$
        and J. David Vergara$^{2}$\thanks{vergara@nuclecu.unam.mx}\\ \\
     $^{1}$   Departamento de F\'{\i}sica \\
        Universidad Aut\'onoma Metropolitana--Iztapalapa\\
       Apartado Postal 55--534, 09340 M\'exico D.F., M\'exico \\  \\
     $^{2}$   Instituto de Ciencias Nucleares \\
        Universidad Nacional Aut\'onoma de M\'exico\\
        Apartado Postal  70-543, 04510 M\'exico D.F., M\'exico \\ \\
      $^{3}$  International Centre for Theoretical Physics,\\
P.O. Box 586, 34100 Trieste, Italy.}
\begin{document}

\date{ May 1996}

\maketitle

\begin{abstract}
We show that the reality conditions to be imposed on Ashtekar variables
to recover  real gravity can be implemented as
second class constraints {\em \`{a} la} Dirac.
Thus, counting gravitational degrees of freedom follows accordingly.
Some constraints of the real theory
turn out to be non-polynomial,  regardless
of the form,
polynomial or non-polynomial,
taken for the reality conditions.
We comment upon the compatibility of
our approach with the recently proposed
Wick transform point of view,  as well as on some alternatives for
dealing with such  second class constraints.

\end{abstract}

\baselineskip=23pt

The formulation of general relativity in terms of the
Ashtekar variables: a
densitized
inverse dreibein ${\tilde e}^{ai}$ together with a complex
$so(3)$ connection
$A_{ai}$ \cite{AA},
has opened up a broad new avenue  of research  \cite{REVAA}.
Reexpressing  general
relativity  as a
theory of a connection has made it  possible  to incorporate
methods which have been very useful and successful in the description of
Yang-Mills  theories.  A most notable
example of this interbreeding  is the recent  application  of
the loop-space
methods
to the quantization of general relativity \cite{LAZOS}. Another
important  advantage  of the
new variables is that the constraints characterizing  general
relativity  turn
out to be polynomial in these
phase space variables.
This simplicity, however, gets afflicted
by the complex character of the new variables: the use of  reality
conditions
is compulsory and the pursue of the Ashtekar program deviates
somewhat from the more
conventional approaches. In fact, the original Ashtekar program
was formulated as an alternative route for classical
general relativity and its canonical quantization, as compared
to the standard  ADM
approach, for example. In the classical theory, general relativity
is considered to be
embbeded in a larger,
complex theory. The restriction to the usual Einstein theory is  
subsequently
made imposing by hand
some reality conditions. In the quantum theory, one first ignores the
reality conditions,
solves the quantum constraints of the complex theory and finally  
imposes the
reality conditions
as  constraints upon the admissible inner products by requiring that
the real
classical observables become self-adjoint operators. The use of  the
reality conditions as a way of selecting an inner product has worked
well in  the case of some models, but
the same has not happened for full gravity so far \cite{KUCHAR}.

The aim of the present work is to reanalyze classical Ashtekar complex
gravity
plus the
reality conditions
in the light of the canonical Dirac approach to  constrained systems. In
particular, we
want to elucidate the status of the reality conditions viewed as Dirac
constraints.
A naive analysis of the standard reality conditions: $q^{ab}$
real and  ${\dot q}{}^{ab}$  real,
as second class constraints,
is  misleading because the count of the number of
degrees of freedom  in the configuration space turns out to
be $-2$ per spatial point, which is obviously  wrong!
A way out  to this situation
has been recently proposed in Ref. \cite{giorgio},
outside the Dirac framework, however,
following a previous result  in Ref. \cite{ART}.

In this work
we start  with
reality conditions, also imposed by hand, but considering them  as
primary constraints
in the
Ashtekar self-dual action. Then,  we systematically apply the Dirac
procedure,
showing that these
reality conditions end up by  being second-class constraints and
recovering the
standard Hamiltonian formulation of general relativity in its
Palatini canonical  form \cite{AA}. In this
way,  we show that  it is possible to extract a satisfactory
canonical
description of general relativity starting from the complex
theory. In fact, we prove that
the original complexified  phase-space, both in the driebein
and in  the
connection, together with
the self-dual Ashtekar action plus  the reality conditions, lead to the
standard phase space of
real Einstein gravity, via the Dirac method.

A closely related issue is that of the polynomiality of the  remaining
first-class
constraints, in terms of the canonical variables. 
It has been shown in  Ref.  
\cite{ABJ}
that the elimination of the associated 
second-class constraints in the complex
Palatini action with self-dual connection  \cite{SM,  JS},  can be  
done without
destroying the polynomial character of the remaining first-class constraints,  
as opposed
to the real Palatini case.  Nevertheless, this nice property
holds only before imposing  the required reality conditions. The  
non-polynomial
choice for such reality conditions made 
in Ref. \cite{ABJ} is briefly discussed there only in  
relation with
the
inner product of the quantum theory and no attempt to  
interpret them
as Dirac constraints is made. We consider  
explicitly  a
non-polynomial form of the reality
conditions and only describe at the end the pertinent
changes produced by  a polynomial  choice.
The  canonical analysis leads in both cases to a  formulation
of  real  general relativity in terms of  non-polynomial constraints.
This feature seems to be compatible
with the recent Wick transform method proposed by Thiemann \cite{tom}
and Ashtekar \cite{wick},
that makes non-polynomiality a `controllable' issue.
Also, this result
seems to be  consistent with the point of view taken in Ref.
\cite{barbero}, where it is shown that  among  a subset of  the
canonical transformations leading to
Ashtekar-type variables in the case of Lorentzian
signature, only the complex  case  leads to a polynomial
Hamiltonian constraint. Thus, real  Ashtekar-type variables
can be used  {\it ab initio}  provided one is willing to accept
a more complicated Hamiltonian constraint.

Let us proceed now to analyze the canonical form of  the self--dual  
action
\cite{ SM, JS}
\begin{equation}
S = \int dtd^3x \left\{ -i\tilde{e}^{ai}\dot{A}_{ai}
   - N {\cal S} - N^a{\cal V}_a -N^i{\cal G}_i\right\},
\label{action}
\end{equation}
where
\begin{equation}
{\cal S}:= \epsilon_{ijk}\tilde{e}^{ai}\tilde{e}^{bj}F_{ab}{}^k,\quad
{\cal V}_a:= \tilde{e}^b{}_j F_{ab}{}^j, \quad
{\cal G}_i:= {\cal D}_a \tilde{e}^a{}_i\,,
\label{const}
\end{equation}
are the constraints of the theory
and $N,N^a,N^i$ are Lagrange multipliers.
The {\em complex} canonical variables are: i)
$\tilde{e}^{ai}:=ee^{ai}$, with $e^{ai}$ being the triad
($e^{ai}e^b{}_i :=q^{ab}$, $q^{ab}$ is the spatial three--metric),
and $a,b,\dots=1,2,3$ are spatial indices, whereas $i,j,\dots=1,2,3$
are so(3) internal indices. Also $e:={\rm det}{\;}e_{bj}$ with $e_{bj}$
being the inverse of $e^{ai}$. ii) $A_{ai}$ is the three--dimensional
projection of the selfdual connection \cite{AA}
with associated covariant derivative ${\cal D}_a\lambda_i=
\partial_a \lambda_i + \epsilon_{ijk} A_a{}^j \lambda^k$
and $F_{ab}{}^i:=
\partial_a A_{b}{}^i - \partial_b A_a{}^i + \epsilon^i{}_{jk}
A_a{}^j A_b{}^k$ is the corresponding
curvature. The above complex  pair of canonically conjugated
variables satisfies
\begin{equation}
\{ \tilde{e}^{ai}(x), A_{bj}(y)\} = i \delta_b{}^a \delta_j{}^i
\delta^3(x,y) \,.
\end{equation}
With these Poisson brackets,  the complex constraints (\ref{const}) are
first class, i.e. for any two of them, say $\cal Q$ and $\cal R$,
we have
\begin{equation}
\{ {\cal Q}, \ {\cal R}  \}_{A, {\tilde e}}\approx 0.\label{CPB}
\end{equation}
Let us remind the reader of  some basic features of the Dirac method
(cf \cite{CLAUDIO}),
which will be used in what follows.
Consider a dynamical system with real
canonical variables $q^i,p_i \quad
i=1,\dots N$ together with a  complete set of real constraints
$G_a(q,p),\ \
a=1,\dots m$. The function $F(q,p)$ is called a first--class
function provided
$\{F,G_a\}= U_a{}^b(q,p) G_b, \forall a$. If this equality does not
hold, $F(q,p)$ is called a second class function. Hereby,
the complete set of constraints gets classified into
first class constraints,
$\Phi_{\alpha},\ \ \alpha=1,\dots,n\leq m$ and second class constraints,
$\chi_A,\ \  A=n+1,\dots m$. The second class set satisfies
$\{\chi_A,\chi_B\}=C_{AB}$, with ${\rm det}\; C_{AB}\neq 0$.
The number of  physical (configuration) degrees of freedom is given by:
$$\rm \frac{1}{2} \left[ dim (phase\, space) - 2 (\#\, of\, 1st\, class\,
constraints) - (\#\, of\, 2nd\,class\,constraints) \right].$$
In this way, the count of real degrees of freedom corresponding to  the
action (\ref{action})
leads to $\frac{1}{2}[ 2\times 2 \times 9 - 2 \times 2\times 7 ] =4$
per spatial point, i.e.
the action (\ref{action}) describes complex gravity. Notice that we have
included an additional
factor of 2 in this count because of the complex character of both the
canonical variables and the constraints. Clearly, one has to
supplement the theory with reality conditions in order to go back to
real gravity.

We shall proceed with the canonical analysis of the action  
(\ref{action}),
together with the necessary reality conditions, by
splitting the complex canonical variables $A_{ai}, \tilde{e}^{bj}$ into
real and imaginary parts according to
\begin{equation}
\tilde{e}^{ai} = E^{ai} + i {\cal E}^{ai}, \quad
A_{bj} = \gamma_{bj} - i K_{bj} \,.
\label{split}
\end{equation}
Most importantly, we will consider all of them (36 real variables)
as configuration variables to begin with. Observe that this
procedure dispenses us from dealing with the complex conjugate of
the Ashtekar variables in the formulation of the reality  
conditions. This is
very convenient because their Poisson brackets would not
even be defined in the theory.

The reality conditions we start with are
\begin{equation}
\psi^{ai}:={\cal E}^{ai} = 0,\quad
\chi_{ai}:=\gamma_{ai} - f_{ai}(E) = 0,
\label{rc1}
\end{equation}
which are subsequently taken as primary  constraints
$\psi^{ai},\chi_{bj}$, supplementing
the action (\ref{action}). We can see that the constraint $\psi^{ai}$
enforces the initially complex triad $\tilde{e}^{ai}$ to be real,
and hence the corresponding three--metric.  Based upon the compatibility
condition between
a real torsion-free connection $\stackrel{0}{ \gamma}_{ai}$ and a
real densitized triad $E^{bj}$
\begin{equation}
D_a E^{bj}= \partial_a E^{bj} + \Gamma_{ac}{}{}^b E^{cj}
+ \epsilon^j{}_{kl} \stackrel{0}{\gamma}_a{}^k E^{bl} - \Gamma_{da}{}{}^d
E^{bj}=0,
\label{comp}
\end{equation}
where  $\Gamma_{ab}{}{}^c$ is the Levi-Civita
connection defined by the metric $h^{ab}:=E^{ai}E^b{}_i$,
we choose the form of $f_{ai}$ as
\begin{equation}
f_{ai}=\frac{1}{2} [ E_{ai}E_c{}^j\epsilon_{jrs} -
2 E_{aj}E_c{}^j\epsilon_{irs} ] E^{dr}\partial_d E^{cs}.
\label{f}
\end{equation}
In this way
we guarantee
that $\gamma_{ai}$ is the real torsion--free connection compatible
with $E^{bj}$, i.e. $\gamma_{ai}=\stackrel{0}{ \gamma}_{ai}$.
Since $E_{ai}$ is the inverse of $E^{bj}$, (\ref{f})
is non-polynomial in $E^{bj}$ and so it is  $\chi_{ai}$.

The definition of the canonical momenta, denoted generically by $\Pi$,
in the action (\ref{action}) leads to the following
constraints
\begin{equation}
\phi_{{\cal E}ai}= \Pi_{{\cal E}ai},\;\;
\phi_{\gamma}{}^{ai} = \Pi_{\gamma}{}^{ai} + i E^{ai},\;\;
\phi_{K}{}^{ai} = \Pi_{K}{}^{ai} + E^{ai},\;\;
\phi_{Eai} = \Pi_{Eai}\,,
\label{mom}
\end{equation}
where we note that $\Pi_{\gamma}{}^{ai}$ is purely imaginary.
The corresponding non-zero Poisson brackets are
$$
\{{\cal E}^{ai}(x), \ \Pi_{{\cal E}bj}(y) \}= \delta^a{}_b
\delta^i{}_j \delta^{(3)}(x,y),\quad
\{{E}^{ai}(x), \ \Pi_{{E}bj}(y) \}= \delta^a{}_b \delta^i{}_j
\delta^{(3)}(x,y) ,
$$
\begin{equation}
\{{\gamma}_{ai}(x), \ \Pi_{\gamma}{}^{bj}(y) \}= \delta_a{}^b
\delta_i{}^j \delta^{(3)}(x,y), \quad
\{{K}_{ai}(x), \ \Pi_{K}{}^{bj}(y) \}= \delta_a{}^b \delta_i{}^j
\delta^{(3)}(x,y) .
\label{PB}
\end{equation}
In this way,  the full set of primary constraints is formed by
(\ref{const}), (\ref{rc1}) and (\ref{mom}), written in terms
of the real variables (\ref{split}). Remarkably, the
analysis of the phase space  could be
also  performed   starting from a real  triad ${\tilde e}^{ai}$.
In this way,
the dimension of the original phase space is   reduced, together with
the number
of required reality conditions. This alternative procedure shows
that there is nothing wrong
in considering a ``mixed phase space": real   ${\tilde e}^{ai}$
and complex  $A_{bj}$, as long as
the canonical analysis is consistently carried out from the very
beginning. Nevertheless, we  choose to deal with both complex
configuration variables because we obtain a more symmetrical
description in doing so.

The next step is to analyze the time evolution of the primary
constraints, together with the classification of the resulting
set of constraints. It turns out to be appropriate to redefine
\begin{equation}
\phi_{Eai}\rightarrow \phi'{}_{Eai}= \phi_{Eai}
+ \alpha_{aibj} \phi_{\gamma}{}^{bj} + \beta_{ai}{}{}^{bj} \chi_{bj}
+ \eta_{aibj} \phi_K{}^{bj},
\label{phip}
\end{equation}
with
\begin{eqnarray}
\alpha_{aibj}(x,y) &=& \frac{\delta f_{bj}(y)}{\delta E^{ai}(x)},\;\; \ \
\beta_{ai}{}{}^{bj}(x,y) = i
\delta_a{}^b\delta_i{}^j\delta^{(3)}(x,y), \nonumber \\
\eta_{aibj}(x,y) - \eta_{bjai}(y,x) &=& i
\left( \frac{\delta f_{ai}(x)}{\delta E^{bj}(y)}-
\frac{\delta f_{bj}(y)}{\delta E^{ai}(x)}  \right),
\label{coef}
\end{eqnarray}
where $\frac{\delta f(y)}{\delta E(x)}$ denotes the corresponding
functional derivative.
Our notation, introduced in Eq.(\ref{phip}), is that the summation
over repeated indices involves also an integral over the three
dimensional space. In this way, for example $\phi_A= \alpha_{AB}
\chi^B$ stands for $\phi_A({\bf x},t)= \int d^3{\bf y} \
\alpha_{AB}({\bf x , y})\chi^B({\bf y}, t)$.

The resulting Poisson brackets matrix for the subset of constraints
(\ref{rc1}), (\ref{mom}) and (\ref{phip}) is
\begin{equation}
\begin{array}{lcccccc}
 & \phi_{\cal E} & \psi & \phi_{\gamma}
 & \chi & \phi_{K} & \phi'{}_E \\ \\
\phi_{\cal E} & 0 & -\delta_a{}^{b}\delta_i{}^{j} \delta^{(3)} & 0  
& 0 & 0 & 0
\\
\psi & \delta_b{}^{a}\delta_j{}^{i} \delta^{(3)} & 0 & 0 & 0 & 0 & 0 \\
\phi_{\gamma} & 0 & 0 & 0 & -\delta_b{}^{a}\delta_j{}^{i}
 \delta^{(3)} & 0 & 0 \\
\chi & 0 & 0 & \delta_a{}^{b}\delta_i{}^{j} \delta^{(3)} & 0 & 0 & 0 \\
\phi_{K} & 0 & 0 & 0 & 0 & 0 & \delta_a{}^{b}\delta_j{}^{i}
 \delta^{(3)}  \\
\phi'{}_E & 0 & 0 & 0 & 0 & -\delta_b{}^{a}\delta_i{}^{j}   
\delta^{(3)}& 0
\end{array},
\label{pbmatrix}
\end{equation}
where $\delta^{(3)}= \delta^{(3)}(x,y)$.
{}From the above matrix we conclude that this subset, which
includes the reality conditions, is second class.
Since  the original constraints (\ref{const}) generate the gauge  
symmetries
of the system,  we would like to keep them as   first--class
constraints. To see whether this is tenable, we start by searching a
redefinition of  them so that they have
zero Poisson brackets with the second--class set.
Notice that the constraints (\ref{const})
only depend on the configuration variables and not
on the momenta. For any of them, say
$\cal R$, we construct
\begin{equation}
{\cal R}'= {\cal R} +
\{ \phi_{{\cal E}bj}, {\cal R} \} \psi^{bj} +
\{ \phi_{\gamma}^{bj}, {\cal R} \} \chi_{bj} +
\{ \phi'{}_{Ebj}, {\cal R} \} \phi_K{}^{bj} -
\{ \phi_{K}{}^{bj}, {\cal R} \} \phi'{}_{Ebj},
\label{rp}
\end{equation}
 which we can verify to fulfill the above requirement.
It remains now to calculate the Poisson brackets among the
constraints of the set ${\cal S}',
{\cal V}'{}_a, {\cal G}'{}_i$. Let us consider
$\{ {\cal Q}'{}, {\cal R}'\}$, i.e. any two of them
\begin{eqnarray}
\{ {\cal Q}'{}, {\cal R}' \} &\approx&
\{ {\cal Q}'{}, {\cal R} \}
= \{ \phi_{{\cal E}ai}, {\cal Q} \} \{ \psi^{ai}, {\cal R} \}
+ \{ \phi_{\gamma}^{ai}, {\cal Q} \} \{ \chi_{ai}, {\cal R} \}  
\nonumber \\
& & +\{ \phi'{}_{Eai}, {\cal Q} \} \{ \phi_K{}^{ai}, {\cal R} \}
- \{ \phi'{}_{Eai}, {\cal R} \} \{ \phi_K{}^{ai}, {\cal Q} \}\,.
\end{eqnarray}
The first two terms yield zero because the constraints $\psi^{ai},  
\chi_{ai}$
have no dependence on the momenta. After substituting the expression
(\ref{phip}) for $\phi'{}_{Eai}$ we are left with
\begin{eqnarray}
\{ {\cal Q}', {\cal R}' \} &\approx&
\{\phi_K{}^{ai}{}, {\cal R} \} \{ \phi_{Eai}, {\cal Q} \}
-\{\phi_K{}^{ai}{}, {\cal Q} \} \{ \phi_{Eai}, {\cal R} \}\nonumber \\
& & + \left(\eta_{aibj}-\eta_{bjai}\right)
    \{ \phi_K{}^{ai}{}, {\cal R} \}
    \{ \phi_K{}^{bj}{}, {\cal Q} \}\nonumber \\
& & + \alpha_{aibj} \left[
\{ \phi_K{}^{ai}{}, {\cal R} \} \{ \phi_{\gamma}{}^{bj}{}, {\cal Q} \}
- \{ \phi_K{}^{ai}{}, {\cal Q} \} \{ \phi_{\gamma}{}^{bj}{}, {\cal R} \}
\right] \,.
\label{qprp}
\end{eqnarray}
A straightforward calculation based on the splitting (\ref{split})
for the first line in (\ref{qprp}) leads to
\begin{eqnarray}
\{\phi_K{}^{ai}{}, {\cal R} \} \{ \phi_{Eai}, {\cal Q} \}
-\{\phi_K{}^{ai}{}, {\cal Q} \} \{ \phi_{Eai}, {\cal R} \}
&=&  -i \left(
\frac{\delta {\cal R}}{\delta A_{ai}}
            \frac{\delta {\cal Q}}{\delta \tilde{e}^{ai}}
           - \frac{\delta {\cal Q}}{\delta A_{ai}}
            \frac{\delta {\cal R}}{\delta \tilde{e}^{ai}}  
\right)\nonumber\\
&=& -i \{ {\cal R}, {\cal Q} \}_{A,\tilde{e}} \approx 0 \,,
\label{aepb}
\end{eqnarray}
according to the first-class property (\ref{CPB}) of
$\cal R$, $\cal Q$ in the original
complex phase space.
Besides, taking into account the corresponding coefficients (\ref{coef})
the second line of (\ref{qprp}) reduces to
$$
i \left( \frac{\delta f_{bj}}{\delta E^{ai}} -
           \frac{\delta f_{ai}}{\delta E^{bj}}\right)
           \frac{\delta{\cal R}}{\delta A_{ai}}
           \frac{\delta{\cal Q}}{\delta A_{bj}},
$$
while the third line is just the negative of this same expression.
We have thus shown that the Poisson brackets among any pair
of the set of constraints
${\cal S}'{}, {\cal V}'{}_a, {\cal G}'{}_i$ is weakly zero.

Summarizing, our system is conveniently
described by the following set of  primary constraints:
$\phi_{{\cal E}ai}$, $\phi_{\gamma}{}^{ai}$,
$\phi_K{}^{ai}$, $\phi'{}_{Eai}$, $\psi^{ai}$, $\chi_{ai}$,
${\cal S}'$, ${\cal V}'{}_a$, ${\cal G}'{}_i$.
Next, we impose the conservation in time
of this whole set, using the total hamiltonian density
\begin{eqnarray}
{\cal H}_{\rm Total}({\bf x})&\approx&
\mu_{\cal E}{}^{ai} \phi_{{\cal E}ai} +
\mu_{\gamma ai} \phi_{\gamma}{}^{ai} +
\mu_{Kai} \phi_K{}^{ai} +
\mu_{E}{}^{ai} \phi'{}_{Eai} +
\lambda_{ai} \psi^{ai} +
\omega^{ai} \chi_{ai} \nonumber \\
& & + N {\cal S}' + N^a {\cal V}'{}_a + N^i {\cal G}'{}_i\,,
\end{eqnarray}
where no integration over the three-space is involved.
The properties of the constraints
${\cal S}',{\cal V}'{}_a,{\cal G}'{}_i\,$, together with
Eqs. (\ref{rp}) and (\ref{aepb}), guarantee that
their Poisson brackets with
${H}_{\rm Total}=\int d^3{\bf x} \ {\cal H}_{\rm Total}({\bf x})$
is weakly zero.
Let us consider now the remaining constraints. By virtue of the
Poisson brackets matrix (\ref{pbmatrix})
together with the definition (\ref{rp}), the calculation
of the corresponding time evolution fixes each of the Lagrange  
multipliers
$\mu_{\cal E}{}^{ai}$, $\mu_{\gamma ai}$,
$\mu_{Kai}$, $\mu_{E}{}^{ai}$, $\lambda_{ai}$, $\omega^{ai}$ equal  
to zero.
We conclude   that there are no secondary constraints
and that the set ${\cal S}'{}, {\cal V}'{}_a, {\cal G}'{}_i$
is first class while the set
$\phi_{{\cal E}ai}$, $\phi_{\gamma}{}^{ai}$,
$\phi_K{}^{ai}$, $\phi'{}_{Eai}$, $\psi^{ai}$, $\chi_{ai}$,
which includes the reality conditions,
is second class.

To  make sure  that ultimately we are dealing with real gravity we
still have
to get the correct number of degrees of freedom, using the Dirac
prescription. Also we must recover the specific form of the constraints
characterizing real gravity.  To do so we start by  examining  the  
complex
character of each constraint. By simple inspection of (\ref{rc1})
and (\ref{mom}) we realize that  each of the second class constraints is
either purely
imaginary or purely real. To proceed with the count we
impose them strongly. This leaves us with $2\times 9\times 4-
6\times 9$ = 18 as the dimension of partially reduced real phase space.
The first class constraints remain first class in the
resulting Dirac brackets. As shown below, this first class set contains
seven
constraints each of which is either purely real or purely imaginary.
In this way the final count yields
$\frac{1}{2}\left[18 - 2\times 7\right]= 2$ real configuration degrees
of freedom per spatial point.

Next we verify our  statement regarding the reality properties of the
first  class constraints and get their explicit  form.   Consider
the Gauss constraint in the partially reduced phase space,  to  
begin with.
It becomes
\begin{equation}
{\cal G}^i\approx {\cal D}_a E^{ai} = \partial_a E^{ai} +
\epsilon^i{}_{jk} \gamma_a{}^{j} E^{ak}- i\epsilon^i{}_{jk}
K_a{}^j E^{ak} \approx -i\epsilon^i{}_{jk} K_a{}^j E^{ak},
\label{gpi}
\end{equation}
where  we have  used the compatibility
condition (\ref{comp}). Hence ${\cal G}^i$ is purely imaginary.
For the vector constraint we have
\begin{eqnarray}
{\cal V}_a\approx E^b{}_i F_{ab}{}{}^i &=&
E^b{}_i R_{ab}{}{}^i - \epsilon^i{}_{jk} E^b{}_i K_b{}^j K_{aj}
-i D_a (K_b{}^i E^b{}_i )  +  i D_b (K_a{}^i E^b{}_i) \nonumber \\
&\approx& i \left[ D_b ( K_a{}^i E^b{}_i - \delta_a{}^b K_c{}^i E^c{}_i)
            \right]
\end{eqnarray}
with $R_{ab}{}{}^i= 2\partial_{[a}\gamma_{b]}{}{}^i + \epsilon^i{}_{jk}
\gamma_a{}^j \gamma_b{}^k$. The last line follows from
the term containing $R$ being zero due to the Bianchi identity and
the term quadratic in $K$ being weakly zero because of (\ref{gpi}).
We see then that the vector constraint is purely imaginary in the   
reduced
phase space.
At last, the scalar constraint can be rewritten as
\begin{equation}
{\cal S} \approx \epsilon_{ijk} E^{ai} E^{bj} F_{ab}{}{}^k
\approx -q R + (E^{al} E^{bm}- E^{bl} E^{am}) K_{bl} K_{am},
\end{equation}
where  use has been made of  the Gauss law (\ref{gpi}) together with  the
compatibility condition (\ref{comp}) to arrive at the final result.
Hereby the scalar constraint is purely real. In each case we have  
recovered the
real  first class constraint corresponding to the Palatini canonical
formulation of  real general relativity, as expected.

Had we started from a real triad together with a complex connection, the
dimension of the original phase space would have been  54. The   
definition of
the momenta
would have produced 27 second class constraints to  which  we  
should add 9
reality conditions.
Taking into account the remaining 7 first class constraints, the  
count of the
degrees of freedom turns out to be correct one
(i.e. two per spatial point), since any of the above constraints is  
either
purely real or purely imaginary.

The analysis performed above was developed using  the  
non-polynomial reality
conditions (\ref{rc1}), (\ref{f}) and it follows that the final
structure of the
complete set of constraints is non-polynomial. One could think that  
changing
the reality conditions to  a polynomial form would solve this issue.
Unfortunately,  the answer is in the negative. For instance, choosing
\begin{equation}
{\tilde \chi}_m{}^a=\epsilon_{mli} E^{bl}\partial_b E^{ai} +
2 \gamma{}_{b[m} E^a{}_{l]} E^{bl},
\label{CRP}
\end{equation}
instead of   the second  equation in (\ref{rc1}), yields a phase space
dependent  matrix (\ref{pbmatrix}) and  non-polynomial
first class constraints (\ref{rp}).

Our results seem to be  compatible with the recently proposed
Wick transform relating real Riemannian and  real Lorentzian
canonical gravity,  which automatically implements the reality
conditions \cite{tom,wick}.
This happens in two respects. First, non-polynomiality
always  shows up at some stage. In our case we face non-polynomiality
when applying  the Dirac method, even if we start from polynomial reality
conditions. On the other hand, in the Wick transform framework
non-polynomiality is manifest in the transform itself.
In the latter case, this issue seems to be
under control, although a  definite  answer has not been attained yet
\cite{wick}.
Second, we were able to  carry over
our analysis to the  Riemannian case,  showing  that the Wick transform
takes the   full set of Riemannian constraints to the Lorentzian ones,
{\em in  } complexified phase space.
This  reinforces the idea that the natural home for such a  
transform is the
complexified
phase space,  as opposed to a space-time interpretation  which  
seems 
improbable \cite{wick}.

We have shown that the reality conditions can be taken as
second class constraints in the Dirac formalism. The standard
treatment calls for the elimination of them in terms of Dirac
brackets. Nevertheless, this procedure would destroy the advantages
of the Ashtekar variables. To surmount this difficulty, we should
look for new directions if we want to keep the  canonical framework.
We can mention two possible strategies along these lines:

i) One can transform the set of second  class constraints to a  
first class
set by adding new variables \cite{CLAUDIO}.
This transformation is rather straightforward
because the
Poisson brackets matrix of the second class constraints
has the  very simple structure  given in (\ref{pbmatrix}). For  
example one
can transform the canonical pair $\psi$, $\phi_{\cal E}$, Eqs.  
(\ref{rc1},
\ref{mom}) in the
following form
\begin{equation}
\widetilde{\phi}_{{\cal E}ai} := \Pi_{{\cal E}ai} +
\widetilde{\Pi}_{{\cal E}ai},
\qquad \widetilde{\psi}^{ai} := {\cal E}^{ai} - \widetilde{{\cal  
E}}^{ai},
\end{equation}
where the new vari\-ables $\widetilde{\Pi}_{{\cal E}ai}$ and
$\widetilde{{\cal E}}^{ai}$ are canon\-i\-cally con\-ju\-gated,
{\it i.e.} \break $\{\widetilde
{\Pi}_{{\cal E}ai}, \widetilde{{\cal E}}^{bj} \} =
\delta_a^b \delta_i^j \delta^{(3)} $. The
new constraints $\widetilde{\phi}_{{\cal E}ai}$ and
$\widetilde{\psi}^{ai}$ have
zero Poisson brackets with the whole set of constraints,
then they are first class. Following the same procedure one can
transform the complete set of second class constraints into a first
class set. Since now all the constraints are first class,
they would define the physical states
once they are promoted to operators.

ii) Another possibility  could be to assume that one of the constraints
in each canonical pair of second class constraints
can be considered as first class,
while the companion is taken to be a gauge fixing  condition.
This procedure would not require the addition of extra variables but its
consistency should
be carefully studied.

Both strategies should directly define the physical states of
real gravity, having the correct number of two configuration
degrees of freedom per spatial point.
Regarding the issue
of either non-polynomiality or, perhaps,
high order  polynomiality, further work has to be done
along these lines before a conclusion can be achieved.
It should be clear that  the present results do not
preclude  accounting for the reality conditions via either the quantum
inner product \cite{AA} or the Wick transform \cite{tom,wick}.

Finally,
we expect that the present canonical analysis will remain valid
for the cases which include
matter couplings as well as space-times with boundaries.

\section*{Acknowledgements}
We thank A. Ashtekar for enlightening discussions regarding this work,
as well as R. Gambini for rearousing our interest in this problem. We also
acknowledge useful conversations with F. Barbero, G. Immirzi and J. Stephany.
The work of HAMT and LFU has been supported  in part by the grant CONACyT
\# 3544-E9311. JDV and LFU also acknowledge the support from grant
UNAM-DGAPA-IN-100694. HAMT would like to thank D.W. Sciama and ICTP-Trieste
for supporting a visit upon which the present work was completed and
SISSA-Trieste for the warm hospitality extended to him.

\end{document}